\documentclass[aps,prl,twocolumn,superscriptaddress,a4paper]{revtex4-1}
\usepackage[dvipdfmx]{graphicx}

\usepackage{amsmath,amsthm,amssymb}
\usepackage{amsmath,bm}
\usepackage{color,ulem}
\usepackage{ifthen}

\newcount \commentout
\newcommand{\1}{\mbox{1}\hspace{-0.25em}\mbox{l}}

\newlength{\figwidth}
\newlength{\figlarge}
\begin{document}
%%%%%%%%%%%%%%%%%%%%%%%%%%%%%%%%%%%%%%%%%%%%%%%%%%%%%%%%%%%%%%%%%%%%%%%
\title{
Non-Hermitian fractional quantum Hall states
}
%%%%%%%%%%%%%%%%%%%%%%%%%%%%%%%%%%%%%%%%%%%%%%%%%%%%%%%%%%%%%%%%%%%%%%%
\author{Tsuneya Yoshida}
\affiliation{Department of Physics, University of Tsukuba, Ibaraki 305-8571, Japan}
\author{Koji Kudo}
\affiliation{Department of Physics, University of Tsukuba, Ibaraki 305-8571, Japan}
\author{Yasuhiro Hatsugai}
\affiliation{Department of Physics, University of Tsukuba, Ibaraki 305-8571, Japan}
%%%%%%%%%%%%%%%%%%%%
%%%%%%%%%%%%%%%%%%%%%%%%%%%%%%%%%%%%%%%%%%%%%%%%%%%%%%%%%%%%%%%%%%%%%%%
\date{\today}
%%%%%%%%%%%%%%%%%%%%%%%%%%%%%%%%%%%%%%%%%%%%%%%%%%%%%%%%%%%%%%%%%%%%%%%
\begin{abstract}
We demonstrate the emergence of a topological ordered phase for non-Hermitian systems.
Specifically, we elucidate that systems with non-Hermitian two-body interactions show a fractional quantum Hall (FQH) state. The non-Hermitian Hamiltonian is considered to be relevant to cold atoms with dissipation.
We conclude the emergence of the non-Hermitian FQH state by the presence of the topological degeneracy and by the many-body Chern number for the ground state multiplet showing $C_{\mathrm{tot}}=1$.
The robust topological degeneracy against non-Hermiticity arises from the many-body translational symmetry.
Furthermore, we discover that the FQH state emerges without any repulsive interactions, which is attributed to a phenomenon reminiscent of the continuous quantum Zeno effect.
\end{abstract}
%%%%%%%%%%%%%%%%%%%%%%%%%%%%%%%%%%%%%%%%%%%%%%%%%%%%%%%%%%%%%%%%%%%%%%%
\pacs{
***
} 
%%%%%%%%%%%%%%%%%%%%%%%%%%%%%%%%%%%%%%%%%%%%%%%%%%%%%%%%%%%%%%%%%%%%%%%
%%%%%%%%%%%%%%%%%%%%%%%%%%%%%%%%%%%%%%%%%%%%%%%%%%%%%%%%%%%%%%%%%%%%%%%
\maketitle
%%%%%%%%%%%%%%%%%%%%%%%%%%%%%%%%%%%%%%%%%%%%%%%%%%%%%%%%%%%%%%%%%%%%%%%

%%%%%%%%%%%%%%%%%%
\paragraph*{Introduction.}---
%%%%%%%%%%%%%%%%%%
In these decades, a variety of novel phenomena have been discovered which arise from topological properties in the bulk~\cite{Thouless_PRL1982,Halperin_PRB82,Hatsugai_PRL93,Kane_Z2TI_PRL05_1,Kane_Z2TI_PRL05_2,Konig_QSHE2007,Qi_TQFTofTI_PRB08,TI_review_Hasan10,TI_review_Qi10,Pesin_TM_NatPhys10,Manmana_TMI1D_PRB12,Yoshida_TMI1D_PRL14,Yoshida_TMI2D_PRB16}.
In particular, the topological ordered phases~\cite{Wen_TopoOrder_SciAdv95} exhibit striking topological phenomena because of correlation effects and the topological properties.
One of the representative examples of topological ordered phases is a fractional quantum Hall (FQH) phase~\cite{Tsui_FQHEExp_PRL82,Laughlin_FQHE_PRL83,Jain_FQHE_PRL89} or a fractional Chern insulator~\cite{Tang_FChern_PRL11,Sun_FChern_PRL11,Neupert_FChern_PRL11,Sheng_FChern_NComm12,Regnalt_FChen_PRX11,Bergholtz_FChern_IntJModPhysB13} which hosts anyons obeying fractional statistics due to the topological degeneracy of the ground states.
The platforms of topological ordered phases extend to bosonic or spin systems. The toric codes for two-~\cite{Kitaev_ToricCode_Elsevier03,Kitaev_KitaevHoneyconmb_Elsevier06} and three-dimensional systems~\cite{Hamma_3DToricCode_PRB05} exemplify the emergence of topological ordered phase in spin systems whose relevance of correlated compounds has been discussed recently~\cite{Takayama_KitaevIr_PRL25,Kasahara_KitaevRuCl_Nature18}.
These topological ordered phases also attract much attention in terms of application to the quantum computations.

Along with the above progress, recent development of technology has pioneered a new type of topological systems, non-Hermitian topological systems.
Extensive studies in these years have discovered various intriguing phenomena described by topological properties of quadratic non-Hermitian matrices.
For instance, it has been elucidated that non-Hermiticity may induce a topological phase which does not have its Hermitian counterpart~\cite{Gong_class_PRX18,Kawabata_gapped_class_arXiv19}. 
Furthermore, non-Hermiticity induces novel gapless excitations in the bulk (e.g., exceptional points~\cite{TKato_EP_book1966,HShen2017_non-Hermi,YXuPRL17_exceptional_ring,VKozii_nH_arXiv17,Yoshida_EP_DMFT_PRB18,Carlstrom_nHknot_arXiv10}, symmetry-protected exceptional rings~\cite{Budich_SPERs_PRB19,Okugawa_SPERs_PRB19,Yoshida_SPERs_PRB19,Zhou_SPERs_Optica19,Kawabata_gapless_class_arXiv19,Yoshida_SPERs_mech_arXiv19,Kimura_SPERs_arXiv19} etc.) which arise from the defectiveness of the Hamiltonian.
In addition, non-Hermiticity may induce a unique bulk-boundary correspondence~\cite{SYao_nHSkin-1D_PRL18,SYao_nHSkin-2D_PRL18,KFlore_nHSkin_PRL18,EElizabet_PRBnHSkinHOTI_PRB19,DBorgnia_arXiv2019} due to the non-Hermitian skin effect.

The above two progresses pose the following crucial question: \textit{what are impacts of non-Hermiticity on topologically ordered phases?}
In particular, it is considered to be significant to elucidate the fate of the topological degeneracy which is source of anyons for Hermitian cases.
In spite of such significant open questions, there are few works addressing non-Hermitian topological ordered phases.

In this paper, we address the above issue, providing a new direction in the study of non-Hermitian topological phases.
Specifically, we demonstrate the emergence of non-Hermitian FQH states in a two-dimensional system with non-Hermitian interactions where spinless fermions are coupled to Abelian gauge fields. This system is considered to be relevant to cold atoms with two-body loss.
We conclude the emergence of non-Hermitian FQH states by combining the following two results: direct computation of the Chern number $C_{\mathrm{tot}}$ indicates that the ground state multiplet is characterized with $C_{\mathrm{tot}}=1$; the topological degeneracy is robust against non-Hermitian interactions, which arises from many-body translational symmetry.
Furthermore, we discover a novel phenomenon for which non-Hermiticity is essential; the FQH state emerges without the repulsive interactions. We find that this intriguing phenomenon arises from interplay between the kinetic term and the dissipative interactions which is reminiscent of the continuous quantum Zeno effect. This unconventional mechanism of the bulk gap may provide a new way to access exotic topological ordered phases.

%%%%%%%%%%%%%%%%%%
\paragraph*{Set up.}---
%%%%%%%%%%%%%%%%%%
The non-Hermitian Hamiltonian analyzed in this paper is shown in Eq.~(\ref{eq: Heff}) which is considered to be relevant to cold atoms.
In order to see this, we start with spinless fermions in Abelian gauge potentials which are decoupled with the environment. After that we take into account the coupling with the environment, yielding a non-Hermitian Hamiltonian.

\begin{figure}[!h]
\begin{minipage}{1\hsize}
\begin{center}
\includegraphics[width=1\hsize,clip]{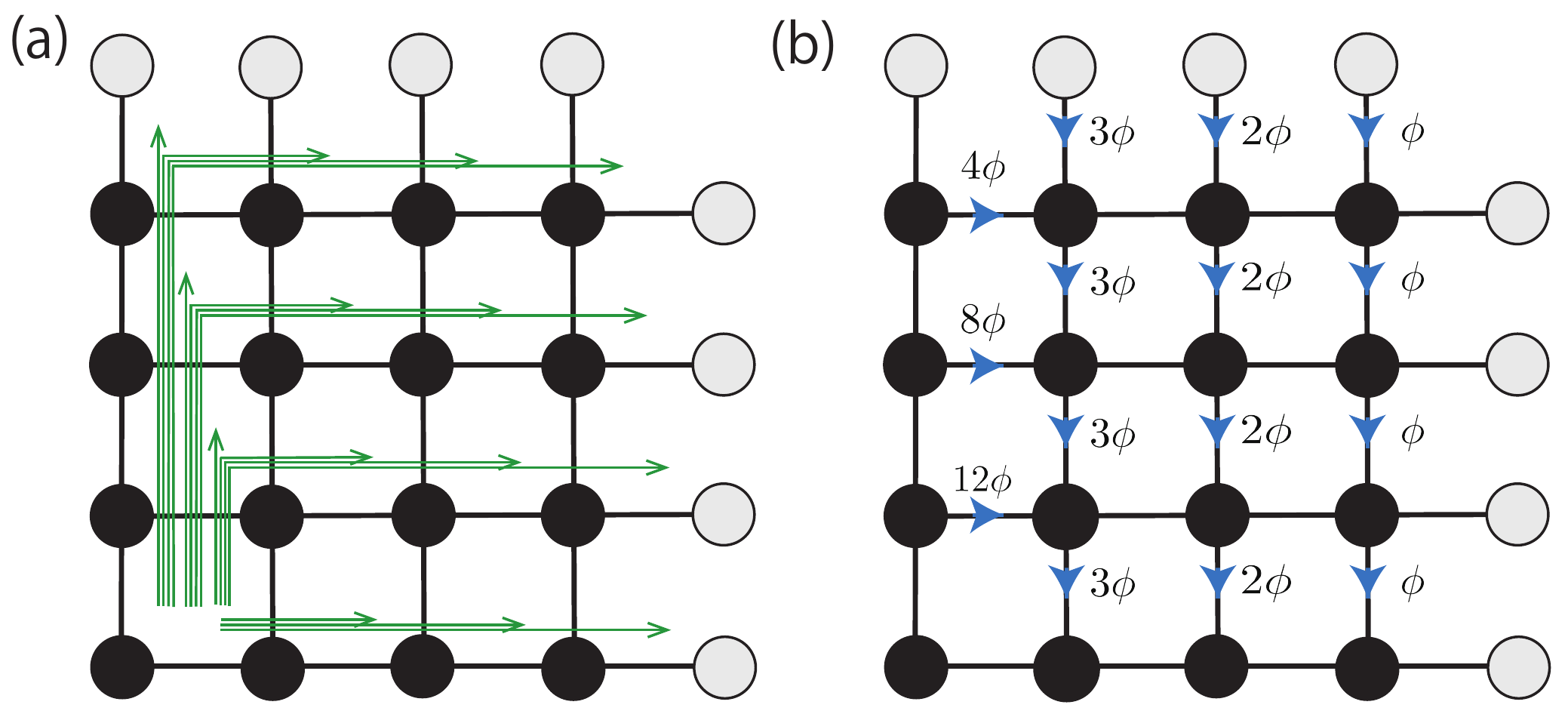}
\end{center}
\end{minipage}
\caption{(Color Online).
Sketch of the model and the string gauge for $N_x=N_y=4$.
We impose the periodic boundary condition for $x$- and $y$-direction.
The green arrows in the right panel represent strings specifying the Peierls phase $\phi_{ij}=2\pi \phi n_{ij}$. Here, $n_{ij}$ denotes the number of string penetrating the bond connecting sites $i$ and $j$, and $\phi$ denotes the flux density $\phi=N_\phi/N_xN_y$.
The right panel illustrates the corresponding Peierls phase; when the fermion hops along the blue arrow, it acquires the phase factor $\phi_{ij}$.
For $\phi$ is multiple of $N^{-1}_x$, the string gauge is reduced to the Landau gauge.
}
\label{fig: model}
\end{figure}

Firstly, we note that the following square lattice system with Abelian gauge fields may be realized for cold atoms
%%%%%%%%%%%%%%%%%%
\begin{eqnarray}
H_0 &=& -\sum_{\langle i,j\rangle} t_0 e^{i\phi_{ij}}c^\dagger_i c_j+V_R\sum_{\langle i,j\rangle} n_in_j.
\end{eqnarray}
%%%%%%%%%%%%%%%%%%
Here, $c^\dagger_i$ creates a spinless fermion at site $i=(i_x, i_y)$ of the two-dimensional system.
If necessary, we rewrite $c^\dagger_i$ as $c^\dagger_{i_xi_y}$.
$t_0$ denotes the hopping between sites. 
The phase factor $\phi_{ij}$ describes the flux penetrating the plaquet. In this paper, we employ the string gauge~\cite{Hatsugai_StringG_PRL99} (see Fig.~\ref{fig: model}).
We define the flux density as $\phi:=N_\phi/N_xN_y$ where $N_\phi$ denotes the number of flux quanta penetrating the $N_x\times N_y$-square lattice.
The filling factor is defined as $\nu:=N_f/N_\phi$ where $N_f$ denotes the number of fermions.
%%%%%%%%%%%%%%%%%%%%%%%%%
%%%%%%%%%%%%%%%%%%%%%%%%%
For fabrication of the above system with cold atoms, the following two ingredients are essential: nontrivial hopping inducing Landau bands and the repulsive interactions. 
The former ones are introduced by rotating the system~\cite{Wilkin_fluxRot_PRL98,Schweikhard_fluxRot_PRL04,NRCooper_fluxRot_AdvPhys08,Furukawa_fluxRotTheory_PRA12} or by optically synthesized gauge fields~\cite{Lin_SynthGauge_Nat09}.
The repulsive interaction ($V_R>0$) may be fabricated by a Feshbach resonance~\cite{Feshbach_FechbachRes_AnnPhys58,Baumann_FeshbachRes_PRA14}.

Now, let us take into account the coupling with the environment.
The time-evolution of such an open quantum system is governed by the Lindblad equation:
%%%%%%%%%%%%%%%%%%
\begin{eqnarray}
\label{eq: Lindbrad_eq}
\partial_t \rho(t) &=& -i[H_0,\rho(t)] -\frac{1}{2}\gamma \sum_{k} \left(L^\dagger_{k} L_{k} \rho(t)  + \rho(t) L^\dagger_k L_k  \right. \nonumber \\
&&\quad \quad\quad \quad\quad \quad\quad \quad\quad \quad \left.-2 L_k\rho L^\dagger_k \right),
\end{eqnarray}
%%%%%%%%%%%%%%%%%%
where $L_k$'s are Lindblad operators describing the loss with the rate $\gamma>0$.
For cold atoms, two-particle loss occurs because of the inelastic scattering~\cite{Scazza_2bdlossExp_NatPhys14,Pagano_2bdloss_PRL15,Hoefer_2bdlossExp_PRL15,Riegger_2bdlossExp_PRL18,Ashida_nHbHubb_PRA16,Nakagawa_nHKondo_PRL18,Yamamoto_nHBCS_arXiv19}, which is described by setting $L_k \to c_ic_{i+\bm{e}_x}$, $c_ic_{i+\bm{e}_y}$. 
Here $\bm{e}_{x(y)}$ denotes the unit vector for each direction, and the lattice constant is set to unity.
When we focus on the short-time evolution, the last term describing the quantum-jump is negligible~\cite{Ashida_nHbHubb_PRA16,Ashida_PTcritical_NatComm17,Nakagawa_nHKondo_PRL18,Yamamoto_nHBCS_arXiv19}.
In this case, we can see that the time-evolution is described by
%%%%%%%%%%%%%%%%%%
\begin{subequations}
\label{eq: Heff}
\begin{eqnarray}
\partial_t \rho(t) &=& -i (H_{\mathrm{eff}}\rho(t) -\rho(t)H^\dagger_{\mathrm{eff}}), \\
 H_{\mathrm{eff}}  &=& H_{\mathrm{kin}} +H_{\mathrm{int}},
\end{eqnarray}
with
\begin{eqnarray}
 H_{\mathrm{kin}}=-\sum_{\langle i,j\rangle} t_0 e^{i\phi_{ij}}c^\dagger_i c_j, && \quad \quad H_{\mathrm{int}}=V\sum_{\langle i,j\rangle} n_in_j.
\end{eqnarray}
\end{subequations}
%%%%%%%%%%%%%%%%%%
We note that the interaction strength takes a complex value; $V=V_R-i\frac{\gamma}{2}$ with $V_R\geq 0$, which makes the Hamiltonian non-Hermitian $H_{\mathrm{eff}}\neq H^\dagger_{\mathrm{eff}}$.

%%%%%%%%%%%%%%%%%%
\paragraph*{Pseudo-potential approach.}---
%%%%%%%%%%%%%%%%%%
Because treating the large size system is numerically difficult, we simplify the problem with calculating the pseudo-potential~\cite{Haldane_FQHEpesudo_PRL83}.
With this approximation, the Hamiltonian is simplified as
%%%%%%%%%%%%%%%%%%
\begin{subequations}
\label{eq: H pesudo}
\begin{eqnarray}
 H'_{\mathrm{eff}}&=& -\sum_{\langle i,j\rangle} t_0 e^{i\phi_{ij}}\tilde{c}^\dagger_i \tilde{c}_j +V\sum_{\langle i,j\rangle} \tilde{c}^\dagger_i \tilde{c}^\dagger_j\tilde{c}_j \tilde{c}_i,
\end{eqnarray}
with
\begin{eqnarray}
H_{\mathrm{kin}}|\phi_\alpha\rangle=|\phi_\alpha\rangle\varepsilon_\alpha, && \quad\quad \tilde{c}^\dagger_i={\sum_\alpha}' \phi^*_{i\alpha} d^\dagger_{\alpha}.
\end{eqnarray}
\end{subequations}
%%%%%%%%%%%%%%%%%%
Here, $|\phi_\alpha\rangle$ denotes the eigenstate of $H_{\mathrm{kin}}$ ($|\phi_\alpha\rangle:=\sum_j \phi_{j\alpha}c^\dagger_{j}|0\rangle$). 
We label the eigenvalues $\varepsilon_\alpha$ so that the relation $\varepsilon_1\leq \varepsilon_2 \leq \cdots$ is satisfied.
$d^\dagger_\alpha$ creates the fermion of the eigenstate $\alpha$.
${\sum_\alpha}'$ denotes the summation over states satisfying $\varepsilon_\alpha \leq \varepsilon_{N_\mathrm{keep}}$; e.g., for $N_{\mathrm{keep}}=N_\phi$ [$N_{\mathrm{keep}}=2N_\phi$], the summation is taken over the lowest Landau levels (LLs) [the lowest and the second lowest LLs]~\cite{LLmixing_footnote}, respectively.

For the numerical computation, we set parameters as $|V|=t_0=1$, $\nu=1/3$, and $N_x=N_y=N$.

%%%%%%%%%%%%%%%%%%
\paragraph*{Hermitian case.}---
%%%%%%%%%%%%%%%%%%
Here we briefly review the results of the Hermitian system ($\mathrm{Im}V=0$) for $\nu=1/3$ where FQH states have been observed.
In this case, three-fold degeneracy is observed for the ground state multiplet which is separated by the bulk gap.
Computing the Chern number $C_{\mathrm{tot}}$ for the ground state multiplet yields $C_{\mathrm{tot}}=1$, which characterizes the topology of the FQH state with the Hall conductance $\sigma_{xy}=1/3$.
For more details, see Sec.~\ref{sec: Hermi FQH app} of Supplemental Material~\cite{supplemental}.

%%%%%%%%%%%%%%%%%%
\paragraph*{Non-Hermitian case.}---
%%%%%%%%%%%%%%%%%%
Now we introduce the imaginary-part $\mathrm{Im}V<0$, which makes the system non-Hermitian. 
Let us start with the definitions of the ground states and the energy gap because the energy spectrum of the non-Hermitian Hamiltonian becomes complex.
We define the ground states with the minimum value of the real-part~\cite{Ashida_nHbHubb_PRA16}.
For our system, these states also have the longest lifetime $\tau$ ($\sim -1/\mathrm{Im}E$ with $\mathrm{Im}E<0$).
Correspondingly, the energy gap is defined as $\Delta=\mathrm{Re}E_e-\mathrm{Re}E_g$ which is natural extension of the Hermitian case. Here, $E_g$ ($E_e$) denotes the energy eigenvalue of the ground state (the first excited state), respectively.
In the following, we numerically show that the FQH state survives even under non-Hermiticity by setting $V=\exp(-i n_\theta\pi/10 )$ with $n_\theta=0,\cdots,5$.

\begin{figure}[!h]
\begin{minipage}{1\hsize}
\begin{center}
\includegraphics[width=0.7\hsize,clip]{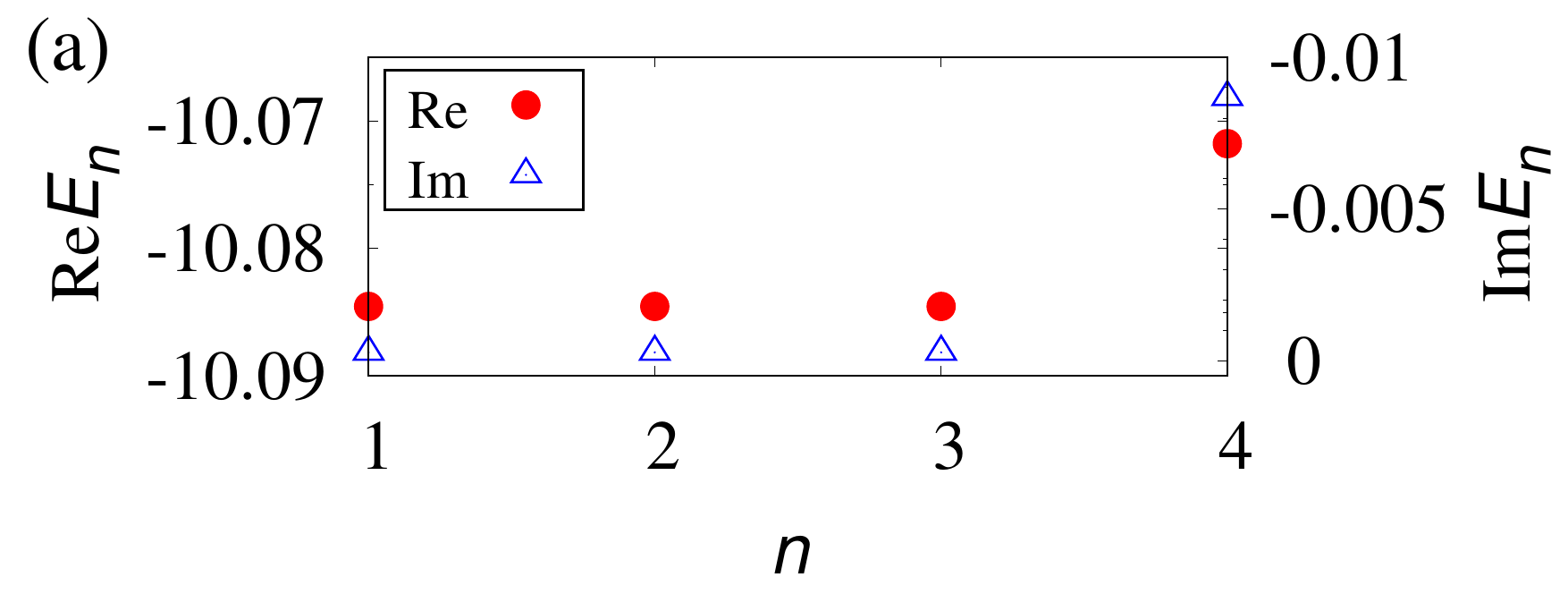}
\end{center}
\end{minipage}
\begin{minipage}{0.45\hsize}
\begin{center}
\includegraphics[width=1\hsize,clip]{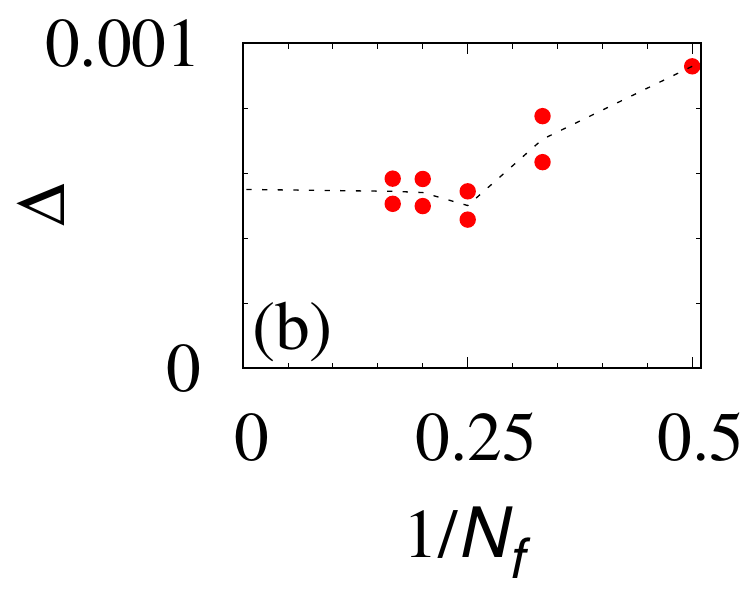}
\end{center}
\end{minipage}
\begin{minipage}{0.45\hsize}
\begin{center}
\includegraphics[width=1\hsize,clip]{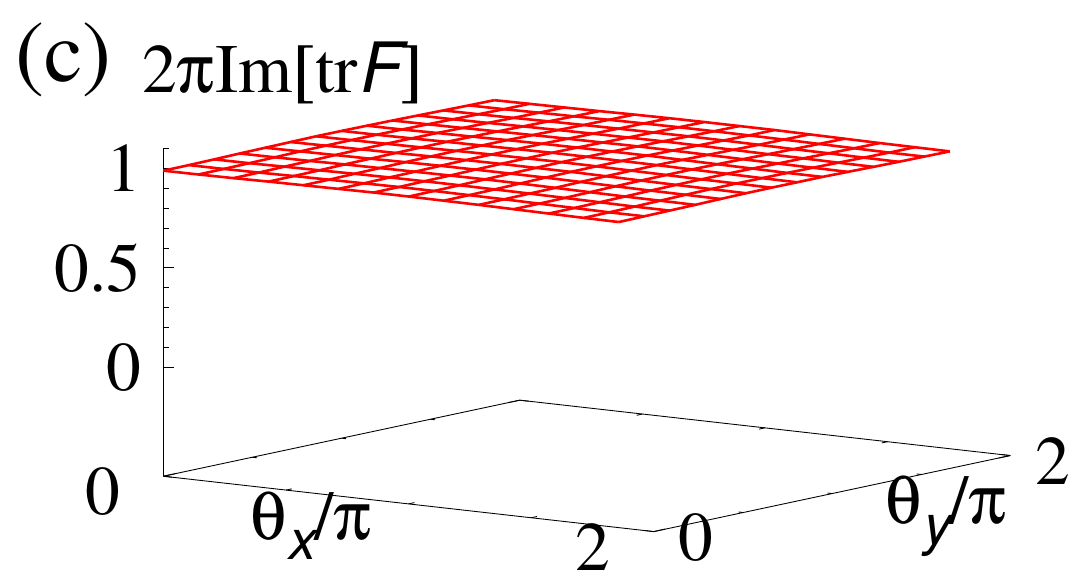}
\end{center}
\end{minipage}
\caption{(Color Online).
Numerical results for $n_\theta=2$ with $V=\exp(-i n_\theta \pi/10)$.
(a) The real- and the imaginary-part of the energy eigenvalues. 
The data are obtained for $n_\theta=2$, $N_\phi=N=9$, and $N_{\mathrm{keep}}=2N_\phi$.
(b) The bulk gap as a function of $N_f$. 
In this plot, the size of the system is chosen so that the flux density $\phi=N_{\phi}/N^2$ satisfies $ 1/45 \leq  \phi \leq 1/40 $. 
The data of panel (b) are obtained for $N_{\mathrm{keep}}=N_\phi$. However, the difference from the gap for $N_{\mathrm{keep}}=2N_\phi$ is less than $\delta \Delta \lesssim 10^{-5}t_0$.
(c) The imaginary-part of the Berry curvature $\mathrm{tr} F$ as a function of $\theta_x$ and $\theta_y$ for $N_{\mathrm{keep}}=N_\phi$. 
For the computation, we divide the two-dimensional space of $\theta$'s into $N_\theta\times N_\theta $-mesh with $N_\theta=14$.
}
\label{fig: datas ithV4piover10}
\end{figure}

As a first step, we focus on the case for $n_\theta=2$. In Fig.~\ref{fig: datas ithV4piover10}(a), we plot the energy spectrum $E_n$ where $n$ labels the states such that $\mathrm{Re}E_1\leq \mathrm{Re}E_2 \leq \cdots$ holds. 
Figure~\ref{fig: datas ithV4piover10}(a) indicates that the three-fold degeneracy can be observed even in the presence of non-Hermitian term. The robustness is attributed to many-body translational symmetry, which we discuss below.
Besides that, in this figure, we can confirm that the lifetime of the ground states is longer than that of excited states.
We note that the energy gap observed in this figure remains finite in the thermodynamic limit, which can be seen in Fig.~\ref{fig: datas ithV4piover10}(b).

%%%%%%%%%%%%%%%%%%%%%%%%%
%%%%%%%%%%%%%%%%%%%%%%%%%

From the above numerical results of the bulk gap and the topological degeneracy, one can expect that the FQH phase survives even in the presence of the non-Hermitian term.
To confirm this, we address the characterization of the topology of the ground states by computing the many-body Chern number for the non-Hermitian case which is defined as follows:
%%%%%%%%%%%%%%%%%%
\begin{subequations}
\label{eq: def NCh}
\begin{eqnarray}
 C_{\mathrm{tot}} &=& \int \frac{d\theta_xd\theta_y}{2\pi i} \mathrm{tr} F(\theta_x,\theta_y), \\
 F_{nm}(\theta_x,\theta_y) &=& \epsilon_{\mu\nu}{}_L\langle \partial_{\mu} \Psi_n | \partial_{\nu} \Psi_m \rangle_{R}. 
\end{eqnarray}
\end{subequations}
%%%%%%%%%%%%%%%%%%
Here, we have imposed the twisted boundary condition: $c^\dagger_{N_x+1,i_y}=e^{i\theta_x} c^\dagger_{ 1,i_y}$ and $c^\dagger_{i_x,N_y+1}=e^{i\theta_y} c^\dagger_{i_x,1}$.
The integral is taken over $ 0 \leq \theta_{x(y)} <2\pi$, respectively.
$F(\theta_x,\theta_y)$ denotes the Berry curvature defined by twisting the boundary conditions. 
$\partial_{\mu}:=\partial/\partial \theta_\mu$. $\epsilon_{\mu\nu}$ ($\mu,\nu=x,y$) is an anti-symmetric matrix with $\epsilon_{xy}=1$.
$| \Psi_n \rangle_{R}$ and ${}_L\langle \Psi_n |$ denote ground states with $n=1,2,3$. The former (latter) ones are right (left) eigenvectors.
The summation is taken over repeated indices.
We note that the Chern number defined above takes integer (for the proof, see Sec.~\ref{sec: proof real C app} of Supplemental Material~\cite{supplemental}).
This fact indicates that the only imaginary-part of the Berry curvature contributes to the Chern number $C_{\mathrm{tot}}$.
Employing the method introduced in Refs.~\cite{Fukui_FHS_mechod_JPSJ05,Fukui_FHS_method_JPSJ07}, we obtain $\mathrm{Im}[\mathrm{tr}F]$. 
In Fig.~\ref{fig: datas ithV4piover10}(b), we can see that the integrand $\mathrm{Im}[\mathrm{tr}F]/2\pi$ becomes almost constant. Evaluating the integration, we obtain $C_{\mathrm{tot}}=1$.

The above data of the bulk gap, the ground state degeneracy, and the Chern number suggest that the ground state is topologically identical to the FQH state with $\sigma_{xy}=1/3$ for the Hermitian case.

\begin{figure}[!h]
\begin{minipage}{1\hsize}
\begin{center}
\includegraphics[width=1\hsize,clip]{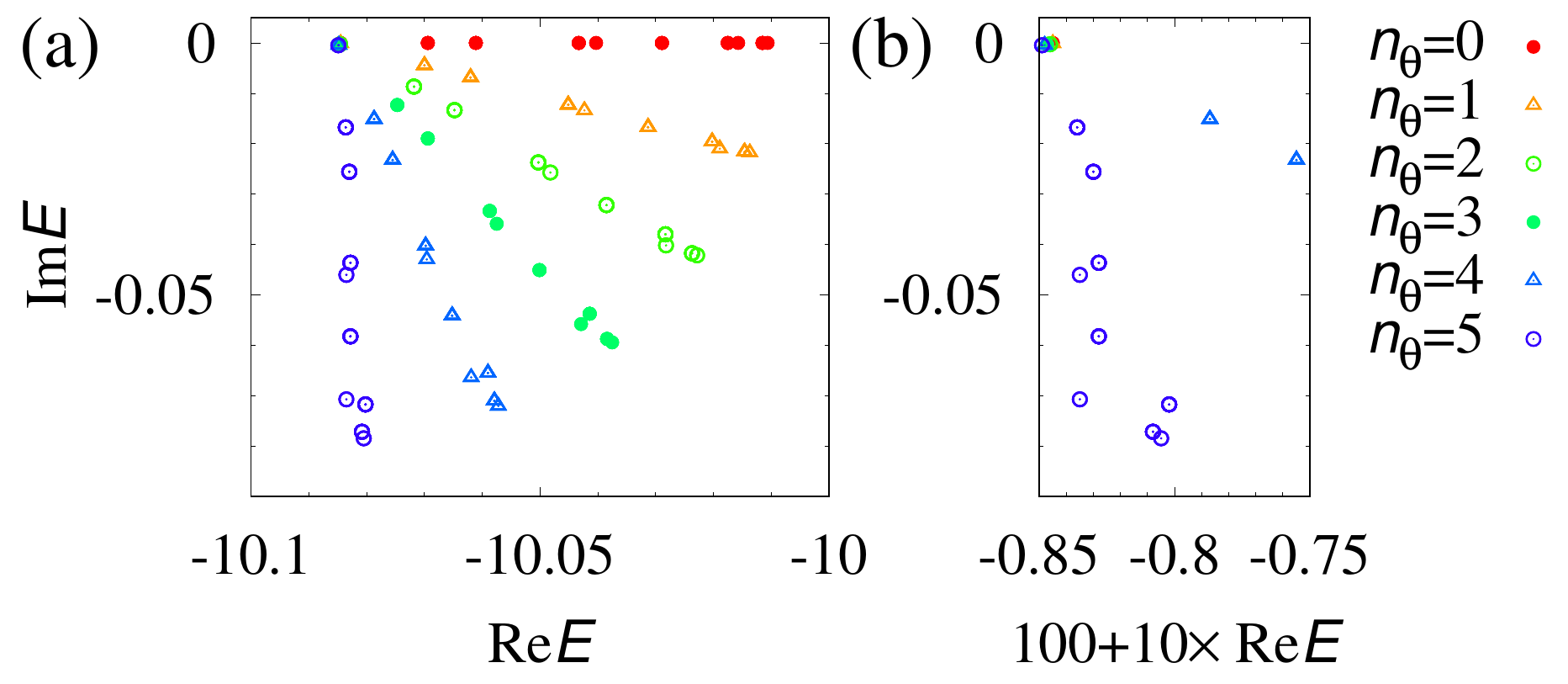}
\end{center}
\end{minipage}
\begin{minipage}{0.7\hsize}
\begin{center}
\includegraphics[width=1\hsize,clip]{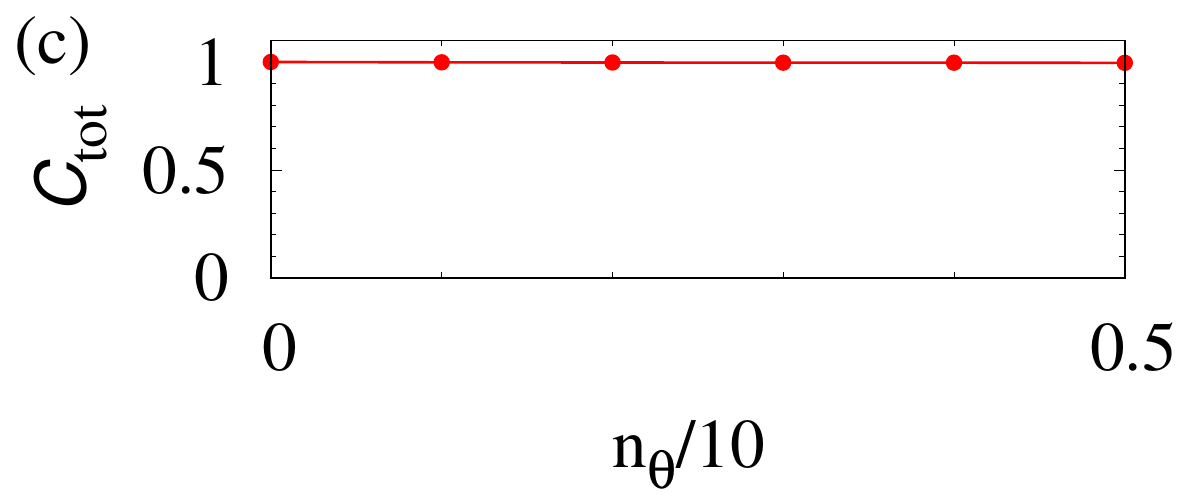}
\end{center}
\end{minipage}
\caption{(Color Online).
(a) Energy spectrum for several values of $n_\theta$ defining interaction with $V=\exp(-i n_\theta\pi/10 )$. Panel (b) shows the magnified data.
(c) Chern number as a function of $n_\theta$.
The data are obtained for $N=9$ and  $N_{\mathrm{keep}}=18$ where both of the lowest and the second lowest LLs are taken into account.
We note that for $n_\theta=5$ the interaction strength $V$ becomes pure imaginary $V=-i$.
}
\label{fig: En_var_ithV}
\end{figure}

In a similar way, we can analyze the system for the other cases of interaction strength.
The results are summarized in Fig.~\ref{fig: En_var_ithV}.
This figure indicates that the FQH state observed for $n_\theta=2$ is adiabatically connected to the one for the Hermitian case; Figures~\ref{fig: En_var_ithV}(a)~and~\ref{fig: En_var_ithV}(b) show that the bulk gap remains finite with decreasing $n_\theta$; Figure~\ref{fig: En_var_ithV}~(c) indicates that the topological properties do not change. 

Intriguingly, Fig.~\ref{fig: En_var_ithV}(b) indicates that the bulk gap opens even for $\mathrm{Re}V=0$, implying the potential presence of the FQH state without the repulsive interaction. 
The details of this issue are addressed below.

%%%%%%%%%%%%%%%%%%%%%%%%%
%%%%%%%%%%%%%%%%%%%%%%%%%

%%%%%%%%%%%%%%%%%%%%%%%%%
\paragraph*{
Robustness of the ground state degeneracy against non-Hermitian interactions.
}---
%%%%%%%%%%%%%%%%%%%%%%%%%
So far, we have numerically observed the three-fold degeneracy of the ground states even in the presence of the non-Hermitian term [see Fig.~\ref{fig: datas ithV4piover10}(a)~and~\ref{fig: En_var_ithV}(a)].
This three-fold degeneracy is due to many-body translational symmetry. Namely, the degeneracy multiple of $\nu^{-1}=2m+1$ ($m\in \mathbb{Z}$) is observed for arbitrary many-body interaction preserving the translational symmetry.
In the following, we discuss the details.

To see this we focus on the case where $N_x=N_y$ and $\phi=n_x/N_x$ ($n_x\in \mathbb{Z}$) holds. Due to the latter condition, the string gauge is reduced to the Landau gauge.
In this case, the kinetic term $H_{\mathrm{kin}}$ preserves the translation symmetry along the $y$-axis.
Namely, the following condition holds; $T_yH_{\mathrm{kin}}T^{-1}_y=H_{\mathrm{kin}}$ where $T_y$ is the translation operator satisfying $T_{y} c^\dagger_{i_xi_y} T^{-1}_{y}=c^\dagger_{i_xi_y+1}$.

Because of the translation symmetry of $H_{\mathrm{kin}}$, one may take the simultaneous eigenstates of $H_{\mathrm{kin}}$ and $T_y$;
%%%%%%%%%%%%%%%%%%
\begin{subequations}
\label{eq: def of |phi k>}
\begin{eqnarray}
H_{\mathrm{kin}} |\varphi_\alpha(k_y)\rangle&=& \varepsilon_\alpha|\varphi_\alpha(k_y)\rangle, \\
T_y |\varphi_\alpha(k_y)\rangle&=& e^{-ik_y}|\varphi_\alpha(k_y)\rangle,
\end{eqnarray}
\end{subequations}
%%%%%%%%%%%%%%%%%%
where $k_y$  denotes the momentum along the $y$-axis ($0 \leq k_y <2\pi$).
Here, let us consider the following gauge transformation: $U_G c^\dagger_{j_xj_y} U^\dagger_G=e^{-i2\pi \phi j_y} c^\dagger_{j_xj_y}$, where $U_G$ is an unitary operator.
Applying the gauge transformation to the eigenstate $|\varphi_\alpha(k_y)\rangle$, we obtain
%%%%%%%%%%%%%%%%%%
\begin{eqnarray}
\label{eq: TU|ky> }
T_y U_G |\varphi_\alpha(k_y)\rangle&=& e^{-i(k_y-2\pi \phi)} U_G|\varphi_\alpha(k_y)\rangle,
\end{eqnarray}
%%%%%%%%%%%%%%%%%%
which means that applying $U_G$ shifts the momentum by $\Delta k_y:=-2\pi \phi$. For the derivation of Eq.~(\ref{eq: TU|ky> }), see Sec.~\ref{sec: TU|ky> app} of Supplemental Material~\cite{supplemental}.

Eq.~(\ref{eq: TU|ky> }) elucidates that the many-body translational symmetry results in the degeneracy multiple of $\nu^{-1}$. This can be seen by noticing the following relation
%%%%%%%%%%%%%%%%%%
\begin{widetext}
\begin{eqnarray}
&& \langle \varphi_{\alpha_1}(k_{y1}), \cdots, \varphi_{\alpha_{N_f}} (k_{yN_f}) |H_{\mathrm{int}} | \varphi_{\beta_1}(k'_{y1}), \cdots, \varphi_{\beta_{N_f}} (k'_{yN_f}) \rangle \\\nonumber
&&= \langle \varphi_{\alpha_1}(k_{y1}), \cdots, \varphi_{\alpha_{N_f}} (k_{yN_f}) | U^\dagger_G H_{\mathrm{int}} U_G| \varphi_{\beta_1}(k'_{y1}), \cdots, \varphi_{\beta_{N_f}} (k'_{yN_f}) \rangle \\\nonumber
&&= \langle \varphi_{\alpha_1}(k_{y1}+\Delta k_y), \cdots, \varphi_{\alpha_{N_f}} (k_{yN_f}+\Delta k_y) | H_{\mathrm{int}} | \varphi_{\beta_1}(k'_{y1}+\Delta k_y), \cdots, \varphi_{\beta_{N_f}} (k'_{yN_f}+\Delta k_y) \rangle,
\end{eqnarray}
\end{widetext}
%%%%%%%%%%%%%%%%%%
which indicates that the matrix element for the subspace labeled by the total momentum $K=\sum_{l} k_{yl}$ equals to the one for the subspace labeled by $K'=K+\Delta k_y N_f$.
Because the shift of the momentum is rewritten as $\Delta k_y N_f=-2\pi\phi N_f=-2\pi\nu$, we can see that the degeneracy of each eigenvalue is multiple of $\nu^{-1}$.

In the above we have seen the relation between the topological degeneracy and the many-body translational symmetry.
In order to support this numerically, we demonstrate that breaking the translational symmetry splits the degeneracy.
Specifically, we compute the energy spectrum in the presence of the following disorder
%%%%%%%%%%%%%%%%%%
\begin{eqnarray}
H_{\mathrm{dis}} &=& \sum_{i} w_i \tilde{c}^\dagger_i\tilde{c}_i,
\end{eqnarray}
%%%%%%%%%%%%%%%%%%
where $w_i$ takes a random value satisfying $ -w_0/2 \leq  w_i \leq w_0/2 $ at each site.
%%%%%%%%%%%%%%%%%%%%%%%%%
\begin{figure}[!h]
\begin{minipage}{0.45\hsize}
\begin{center}
\includegraphics[width=1\hsize,clip]{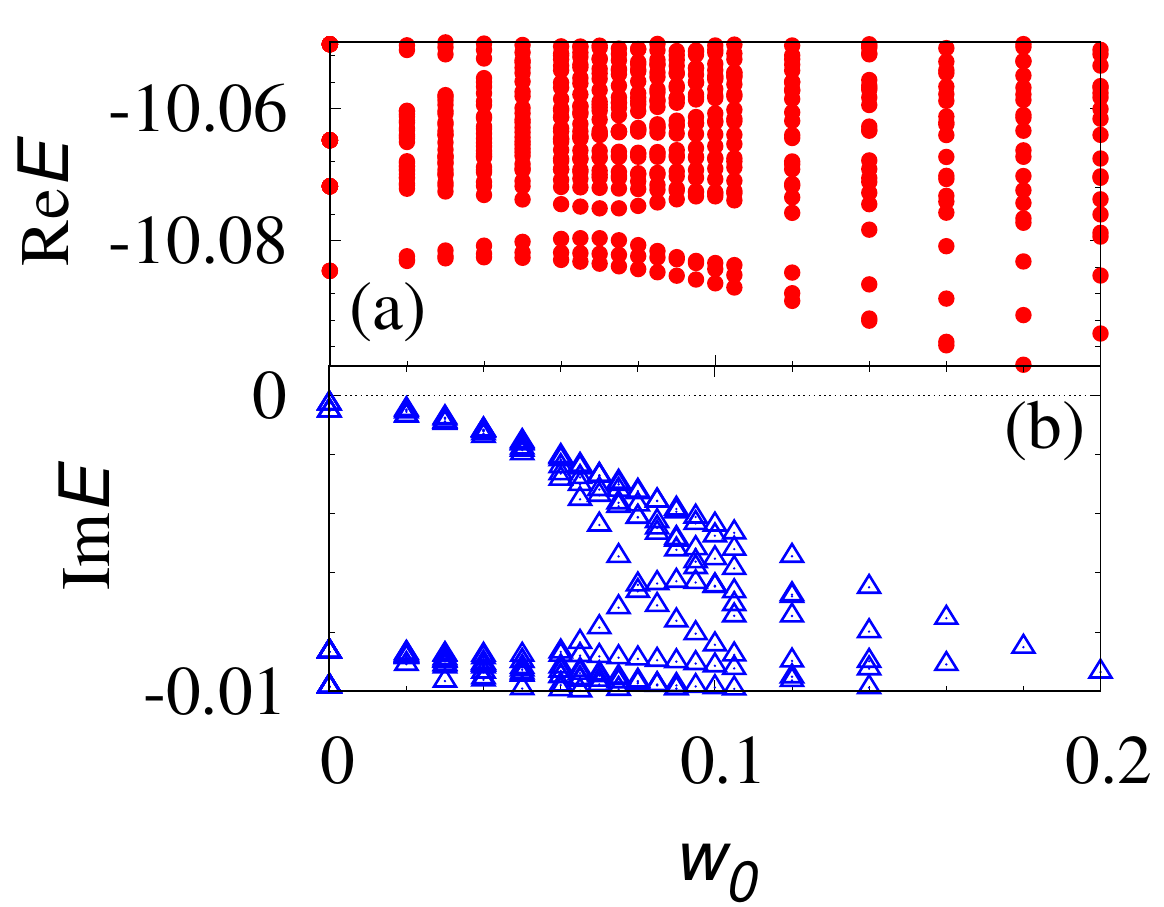}
\end{center}
\end{minipage}
\begin{minipage}{0.45\hsize}
\begin{center}
\includegraphics[width=1\hsize,clip]{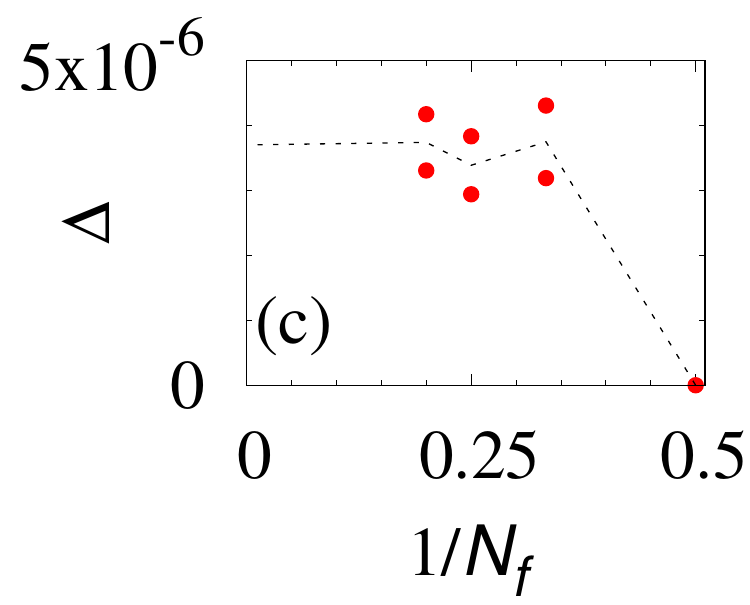}
\end{center}
\end{minipage}
\caption{(Color Online).
(a) [(b)] The real- [imaginary-] part of the energy eigenvalues as functions of disorder strength.
Turning on disorder $w_0$ splits the three-fold degeneracy observed for $w_0$.
The data are obtained for $N_\phi=N=9$ and $n_\theta=2$ with $V=\exp(-i n_\theta \pi/10)$.
(c) The bulk gap as a function of $N_f$ which is obtained for $H_{\mathrm{ptb}}$ [see Eq.~(\ref{eq: ptb_H2nd})].
Energy difference of the ground state multiplet is of the order of $10^{-13}t_0$ which is much smaller than the energy gap.
}
\label{fig: dis and ImV scale}
\end{figure}
%%%%%%%%%%%%%%%%%%%%%%%%%
In Fig.~\ref{fig: dis and ImV scale}(a) [(b)], the real- [imaginary-] part of the energy eigenvalues are plotted against disorder strength $w_0$, respectively.
These figures indicate that breaking the translational symmetry lifts the three-fold degeneracy of the ground states.

The above results indicate that the many-body translational symmetry results in the robustness of topological degeneracy against non-Hermiticity. 
Our numerical data elucidate that the topological degeneracy can be observed for $1/45 \leq  \phi < 1/40$ where the string gauge cannot be reduced to the Landau gauge.

%%%%%%%%%%%%%%%%%%%%%%%%%
\paragraph*{
FQH state without the repulsive interaction.
}---
%%%%%%%%%%%%%%%%%%%%%%%%%
Figures~\ref{fig: En_var_ithV}(b)~and~\ref{fig: En_var_ithV}(c) imply that the FQH state emerge without the repulsive interaction.
In the following, we elucidate the origin of the FQH state for $\mathrm{Re} V=0$.

Firstly, we point out that the origin of the gap is the interplay between the kinetic term $H_{\mathrm{kin}}$ and the non-Hermitian interaction $H_{\mathrm{int}}$ (i.e., the mixing between Landau bands).
Applying the perturbation theory, we obtain the following Hamiltonian acting on the space spanned by the states in the lowest LLs,
%%%%%%%%%%%%%%%%%%
\begin{eqnarray}
\label{eq: ptb_H2nd}
H_{\mathrm{ptb}}= P_0 H_{\mathrm{int}}P_0 +P_0 H_{\mathrm{int}} P_1 \frac{1}{E^0_g-H_{\mathrm{kin}}} P_1 H_{\mathrm{int}} P_0. \nonumber \\
\end{eqnarray}
%%%%%%%%%%%%%%%%%%
Here, $P_n$ denotes the projection operator to the subspace where $n$-fermions are excited to the second lowest LLs. $E^{0}_g$ is the ground state energy for $V=0$.
We have omitted the constant term arising from $P_0 H_{\mathrm{kin}} P_0$.
Noticing that the prefactor of the last term is $(\mathrm{Im}V)^2/\hbar \omega_0$, we can see that the last term serves as the repulsive interaction.
Here $\hbar \omega_0$ denotes the energy gap between the lowest LLs and the second lowest LLs for $V=0$.

Diagonalizing the effective Hamiltonian~(\ref{eq: ptb_H2nd}), we plot the energy gap $\Delta$ as a function of $N_f$ in Fig.~\ref{fig: dis and ImV scale}(c). This figure indicates that the energy gap remains finite in the thermodynamic limit. We also note that the three-fold degeneracy of the ground states is also observed.
Thus, one may consider that the gapped state is the FQH state, which is confirmed by the numerical computation yielding $C_{\mathrm{tot}}=1$ [see Fig.~\ref{fig: En_var_ithV}(c)].

Therefore, we conclude that the FQH state emerges without the repulsive interaction ($\mathrm{Re}V=0$) which is adiabatically connected to the FQH state with $\sigma_{xy}=1/3$ for the Hermitian case.

We stress that the bulk gap opens due to two-body loss inducing the effective repulsive interaction~\cite{ReV_prefac_footnote}, which is reminiscent of the continuous quantum Zeno effect~\cite{Syassen_Zeno_Science08,Mark_Zeno_PRL12,Barontini_Zeno_PRL13,Zhu_Zeno_PRL14,Tomita_Zeno_SciAdv17,Ashida_nHbHubb_PRA16,Nakagawa_nHKondo_PRL18,Yamamoto_nHBCS_arXiv19}.

%%%%%%%%%%%%%%%%%%
\paragraph*{
Summary and outlook.
}---
%%%%%%%%%%%%%%%%%%
In this paper, by focusing on the FQH system at $\nu=1/3$, we have analyzed impact{s} of non-Hermiticity on topological ordered phases.
We have elucidated the robustness of topological degeneracy against non-Hermitian interactions which arises from many-body translational symmetry.
Combining the numerical results of the Chern number $C_{\mathrm{tot}}=1$ and the topological degeneracy leads us the conclusion that non-Hermitian Hamiltonian~(\ref{eq: Heff}b) shows the FQH state.
Furthermore, we have discovered that the FQH state emerges without repulsive interactions ($\mathrm{Re}V=0$).
This intriguing behavior arises from the effective repulsion induced by the two-body loss, which is reminiscent of the continuous quantum Zeno effect.

We finish this article with comments on future directions.
In Fig.~\ref{fig: datas ithV4piover10}(d), we have numerically observed that the Berry curvature is almost independent of $\theta$'s, which implies that the computation of the Chern number may be simplified by defining the non-Hermitian counterpart of the one-plaquet Chern number for the Hermitian case~\cite{Hastings_PlaqCh_CommMath15,Koma_PlaqCh_arXiv15,Watanabe_PRB18,Kudo_PlaqCh_PRL19}. We leave the extension of one-plaquet Chern number to non-Hermitian systems as a future work.
In addition, we have observed that the interplay between the dissipative two-body interaction and the kinetic term yields four-body interactions which open the bulk gap and yield the FQH state. 
This unconventional mechanism of gap opening may provide new direction to access exotic topological ordered states induced by many-body interactions higher than two-body (e.g., the Moore-Read state~\cite{Moore_MRstate_NuclPhys91,Greiter_MRstate_PRL91}).
Hunting such exotic topological ordered states is also left as a significant issue to be addressed.

%%%%%%%%%%%%%%%%%%%%%%%%
\paragraph*{
Acknowledgement.
}---
%%%%%%%%%%%%%%%%%%%%%%%%
This work is partly supported by JSPS KAKENHI Grants 
No.~JP16K13845, %YH
No.~JP17H06138, %YH
and No.~JP18H05842. %TY
A part of numerical calculations were performed on the supercomputer at the ISSP in the University of Tokyo.

%%%%%%%%%%%%%%%%%%%%%%%%%%%%%%
%\bibliography{nHFQHE.bib}

\begin{thebibliography}{85}
\bibitem [{LLm()}]{LLmixing_footnote}%
  \BibitemOpen
  \href@noop {} {}\bibinfo {note} {{ For $N_{\mathrm{keep}}=2N_\phi$, we
  can take into account effects of Landau band mixing which induces an
  intriguing phenomena as we see below }}\BibitemShut {NoStop}%
\bibitem [{ReV()}]{ReV_prefac_footnote}%
  \BibitemOpen
  \href@noop {} {}\bibinfo {note} {{ 
  %
  For $V=V_R>0$ the prefactor of the second term of Eq.~(\ref{eq: ptb_H2nd}) has the opposite sign [$-(\mathrm{Re}V)^2/\hbar \omega_0 $]. We have numerically confirmed that for $V=V_R>0$, the second order term suppresses the bulk gap induced by the first term of Eq.~(\ref{eq: ptb_H2nd})
   %
  }}\BibitemShut {NoStop}%
\end{thebibliography}

\clearpage

\renewcommand{\thesection}{S\arabic{section}}
\renewcommand{\theequation}{S\arabic{equation}}
\setcounter{equation}{0}
\renewcommand{\thefigure}{S\arabic{figure}}
\setcounter{figure}{0}
\renewcommand{\thetable}{S\arabic{table}}
\setcounter{table}{0}
\makeatletter
\c@secnumdepth = 2
\makeatother

\onecolumngrid
\begin{center}
 {\large 
  \textmd{{\bfseries Supplemental Materials}}
 }
% {\large \textmd{Supplemental Materials} 
% \\[0.3em]
% {\bfseries Non-Hermitian fractional quantum Hall states}
% }
\end{center}

\setcounter{page}{1}

%\appendix
%%%%%%%%%%%%%%%%%%
\section{
Detailed results of the Hermitian case
}
\label{sec: Hermi FQH app}
%%%%%%%%%%%%%%%%%%

We here summarize the results of the Hermitian case~\cite{Laughlin_FQHE_PRL83,Niu_HallCond_PRB85,Haldane_TopoDeg_PRL85,Sheng_Chern_PRL03,Tang_FChern_PRL11,Sun_FChern_PRL11,Neupert_FChern_PRL11,Sheng_FChern_NComm12,Regnalt_FChen_PRX11,Bergholtz_FChern_IntJModPhysB13,Kudo_FQHE_JPSJ17}. 
It is well-known that the nearest neighbor interaction ($V>0$) opens the bulk gap, separating exited states and the ground states whose topological degeneracy is three.
Extrapolating the obtained bulk gap for each value of $N_f$, we can confirm that the bulk gap remains finite in the thermodynamic limit. 
We can also numerically confirm the three-fold degeneracy for the ground state multiplet~\cite{Haldane_TopoDeg_PRL85}.
%%%%%%%%%%%%%%%%%%%%%%%%%
\begin{figure}[!h]
\begin{minipage}{0.45\hsize}
\begin{center}
\includegraphics[width=1\hsize,clip]{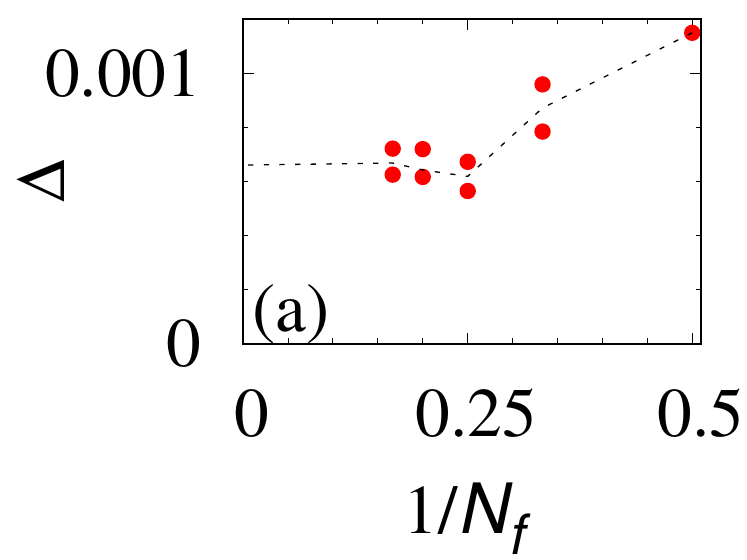}
\end{center}
\end{minipage}
\begin{minipage}{0.45\hsize}
\begin{center}
\includegraphics[width=1\hsize,clip]{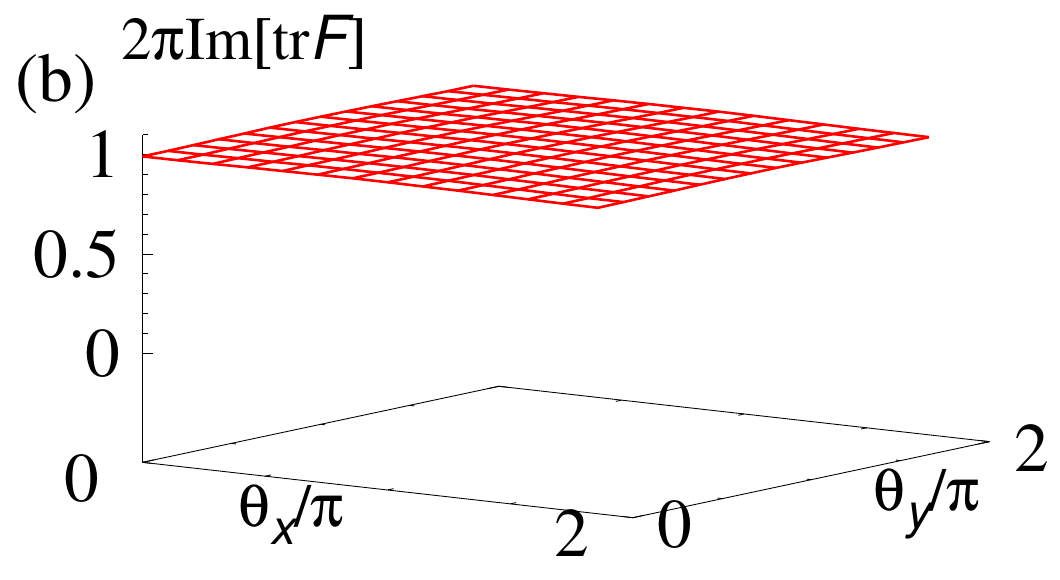}
\end{center}
\end{minipage}
\caption{(Color Online).
(a) The bulk gap as a function of $N_f$ for $N_{\mathrm{keep}}=N_\phi$. These data are obtained in a similar way as Fig.~\ref{fig: datas ithV4piover10}(a) in the main text.
(b) Berry curvature $\mathrm{tr} F/2\pi i$ as a function of $\theta_x$ and $\theta_y$.
}
\label{fig: datas Hermi FQHE}
\end{figure}
%%%%%%%%%%%%%%%%%%%%%%%%%
The topological property of the gapped state can be characterized by many-body Chern number $C_{\mathrm{tot}}$ with twisting the boundary condition.
Computing the Chern number for the ground state multiplet yields $C_{\mathrm{tot}}=1$~\cite{Sheng_Chern_PRL03}.

The above numerical data indicate that the nearest neighbor interaction $V>0$ resluts in the FQH state with $\sigma_{xy}=1/3$.

%%%%%%%%%%%%%%%%%%
\section{
Chern number and Berry connection
}
\label{sec: proof real C app}
%%%%%%%%%%%%%%%%%%
Here, we show that the Chern number defined in Eq.~(\ref{eq: def NCh}b) takes an integer.

Now, consider a two-dimensional parameter space $(\theta_x,\theta_y)$ with $0\leq \theta_{x(y)} <2\pi$. 
Then, we divide the two-dimensional space into two regions, I and II since taking the unique gauge may not be allowed.
Because both of the gauges are available on the boundary of the region I and II, the eigenvectors are related to each other with an invertible matrix $M$;
%%%%%%%%%%%%%%%%%%
\begin{subequations}
\label{eq: gauge app}
\begin{eqnarray}
|\Psi^{II}_n\rangle_R&:=& |\Psi^I_{n'}\rangle_R M_{n'n}, \\
{}_L\langle \Psi^{II}_n |&:=& M^{-1}_{nn'}{}_L\langle \Psi^I_{n'} |,
\end{eqnarray}
\end{subequations}
%%%%%%%%%%%%%%%%%%
where the summation is taken over repeated indices.

Let us evaluate the integration of Eq.~(\ref{eq: def NCh}b).
Applying Stokes' theorem, we can rewrite it as
%%%%%%%%%%%%%%%%%%
\begin{subequations}
\label{eq: C = int A}
\begin{eqnarray}
C_{\mathrm{tot}} &=& \frac{1}{2\pi i}\int_C d\bm{\theta} \cdot \left( \mathrm{tr}\bm{A}^I-\mathrm{tr}\bm{A}^{II}\right),
\end{eqnarray}
with 
\begin{eqnarray}
\bm{A}^\alpha_{nm} &=& {}_L\langle \Psi^\alpha_n | \bm{\nabla} \Psi^\alpha_m\rangle_R.
\end{eqnarray}
\end{subequations}
%%%%%%%%%%%%%%%%%%
Here, the integral of Eq.~(\ref{eq: C = int A}a) is taken along the boundary. $d\bm{\theta}:=(d\theta_x,d\theta_y)$, and $\bm{\nabla}:=(\partial/\partial\theta_x,\partial/\partial\theta_y)$.

Eq.~(\ref{eq: C = int A}a) can be further simplified as follows:
%%%%%%%%%%%%%%%%%%
\begin{eqnarray}
C_{\mathrm{tot}} &=& \frac{1}{2\pi i}\int_C d\bm{\theta} \cdot \mathrm{tr}\left(M^{-1}\bm{\nabla}M \right) \nonumber \\
                 &=& \frac{1}{2\pi i}\int_C d\bm{\theta} \cdot \bm{\nabla} \mathrm{tr} \log M \nonumber \\
                 &=& \frac{1}{2\pi i}\int_C d\bm{\theta} \cdot \bm{\nabla} \log \mathrm{det} M. \nonumber \\
\end{eqnarray}
%%%%%%%%%%%%%%%%%%
Noticing that $\mathrm{det}M$ is a single-valued function, we can see that the integral is reduced to the winding number 
%%%%%%%%%%%%%%%%%%
\begin{eqnarray}
C_{\mathrm{tot}} &=& \frac{1}{2\pi} \mathrm{Im} \int_C d\bm{\theta} \cdot \bm{\nabla} \log \mathrm{det} M,\nonumber \\
                 &\in& \mathbb{Z}.
\end{eqnarray}
%%%%%%%%%%%%%%%%%%
Therefore, we can conclude that the Chern number takes integer.

%%%%%%%%%%%%%%%%%%
\section{
Derivation of Eq.~(\ref{eq: TU|ky> })
}
\label{sec: TU|ky> app}
%%%%%%%%%%%%%%%%%%
As mentioned in the main text, Eq.~(\ref{eq: TU|ky> }) can be obtained for the Landau gauge with $N_x=N_y$.

As a preparation, we discuss the translational symmetry in term of the eigenvectors~(\ref{eq: H pesudo}b). %   (\ref{eq: Heff}b).
The state $|\varphi_\alpha(k_y)\rangle$ can be expanded as
%%%%%%%%%%%%%%%%%%
\begin{eqnarray}
|\varphi_\alpha(k_y)\rangle &=& \sum_{i_x,i_y}\varphi_{i_xi_y\alpha} c^\dagger_{i_xi_y} |0\rangle.
\end{eqnarray}
%%%%%%%%%%%%%%%%%%
Because $|\varphi_\alpha(k_y)\rangle$ is an eigenstate of $T_y$ [see Eq.~(\ref{eq: def of |phi k>})], we have
%%%%%%%%%%%%%%%%%%
\begin{eqnarray}
\label{eq: trans of phi_ja}
 \varphi_{i_xi_y-1\alpha} &=& e^{-ik_y}\varphi_{i_xi_y\alpha}.
\end{eqnarray}
%%%%%%%%%%%%%%%%%%

Now we show that Eq.~(\ref{eq: TU|ky> }) holds.
This can be seen by analysing whether $U_G|\varphi_\alpha(k_y)\rangle$ is an eigenstate of $T_y$;
%%%%%%%%%%%%%%%%%%
\begin{widetext}
\begin{eqnarray}
T_y U_G |\varphi_\alpha(k_y)\rangle 
&&= \sum^{N_x}_{j_x=1} \sum^{N_y}_{j_y=1} e^{-i2\pi\phi j_y} T_y \varphi_{j_xj_y\alpha} c^\dagger_{j_xj_y} |0\rangle \nonumber \\
&&= \sum^{N_x}_{j_x=1} \sum^{N_y}_{j_y=1} e^{-i2\pi\phi j_y} \varphi_{j_xj_y\alpha} c^\dagger_{j_xj_y+1} |0\rangle \nonumber \\
&&= \sum^{N_x}_{j_x=1} \sum^{N_y+1}_{j_y=2} e^{-i2\pi\phi (j_y-1)} \varphi_{j_x(j_y-1)\alpha} c^\dagger_{j_xj_y} |0\rangle \nonumber \\
&&= \sum^{N_x}_{j_x=1} \sum^{N_y+1}_{j_y=2} e^{-i2\pi\phi (j_y-1)-ik_y} \varphi_{j_xj_y\alpha} c^\dagger_{j_xj_y} |0\rangle \nonumber \\
&&= \sum^{N_x}_{j_x=1} e^{-ik_y} \left[ \sum^{N_y}_{j_y=1} e^{-i2\pi\phi (j_y-1)} \varphi_{j_xj_y\alpha} c^\dagger_{j_xj_y} |0\rangle +\left( e^{-i2\pi\phi N_y}-1 \right) \varphi_{j_x1\alpha} c^\dagger_{j_x1}|0\rangle \right] \nonumber \\
&&= e^{-i(k_y-2\pi\phi)}U_G |\varphi_\alpha(k_y)\rangle.
\end{eqnarray}
\end{widetext}
%%%%%%%%%%%%%%%%%%
Here, from the third to the fourth line we have used Eq.~(\ref{eq: trans of phi_ja}).
From fifth to the last line, we have used the relation $\phi N_y=1$ which is satisfied for the system with the Landau gauge and for $N_x=N_y$.

\end{document}